\documentclass[aps,prb,twocolumn,showpacs,superscriptaddress]{revtex4}
\usepackage{amssymb}
\usepackage{graphicx}
\usepackage{epsfig}
\usepackage{amsmath}
\usepackage{amsfonts}
\usepackage{bm}
\newcommand{\ox}{$^{17}$O}
\newcommand{\va}{$^{51}$V}
\newcommand {\cavo} {CaV$_2$O$_4$}

\begin{document}

\title{$^{17}$O and $^{51}$V NMR for the zigzag spin-1 chain compound CaV$_2$O$_4$}
\date{\today}
\author{X. Zong}
\affiliation{Ames Laboratory, Ames, Iowa 50011}
\affiliation{Department of Physics and Astronomy, Iowa State University, Ames, Iowa 50011}
\author{B. J. Suh} 
\altaffiliation[]{Permanent address: Department of Physics, the Catholic University of Korea,
 Bucheon 420-743, Korea;  e-mail: bjsuh@catholic.ac.kr}
\author{A. Niazi}
\author{J. Q. Yan}
\affiliation{Ames Laboratory, Ames, Iowa 50011}
\author{D. L. Schlagel}
\author{T. A. Lograsso}
\affiliation{Materials $\&$ Engineering Physics Program, Ames Laboratory, Ames, Iowa 50011}
\author{D. C. Johnston}
\affiliation{Ames Laboratory, Ames, Iowa 50011}
\affiliation{Department of Physics and Astronomy, Iowa State University, Ames, Iowa 50011}

\begin{abstract}
     \va\ NMR studies on CaV$_2$O$_4$ single crystals and \ox\ NMR studies on \ox-enriched  powder samples are reported.  The temperature dependences of the \ox\ NMR line width and nuclear spin-lattice relaxation rate give strong evidence for a long-range antiferromagnetic transition at $T_{\rm N}=78$~K in the powder. Magnetic susceptibility measurements show that $T_{\rm N}=69$~K in the crystals. A zero-field \va\ NMR signal was observed at low temperatures ($f \approx 237$~MHz at 4.2~K) in the crystals. The field swept spectra with the field in different directions suggest the presence of two antiferromagnetic  substructures.  Each substructure is collinear, with the easy axes of the two substructures separated by an angle of $19(1)^\circ$, and with their average direction pointing approximately along the $b$-axis of the crystal structure.  The two spin substructures contain equal number of spins.  The temperature dependence of the ordered moment, measured up to 45~K, shows the presence of an energy gap $E_{\rm G}$ in the antiferromagnetic spin wave excitation spectrum. Antiferromagnetic spin wave theory suggests that $E_{\rm G}/k_{\rm B}$ lies between 64 and 98~K\@.  
   
\end{abstract}
\maketitle

\section{Introduction}

   Frustrated magnetic systems have attracted a lot of research interest because such systems often exhibit interesting low temperature properties.\cite{Diep} The zig-zag spin chain with antiferromagnetic interactions between nearest- and next-nearest-neighbors is about the most simple frustrated system. In a zig-zag spin chain system with spin $S=1$, the ground state phase diagram (at temperature $T=0$) as a function of XXZ anisotropy and ratio between nearest-neighbor (NN) and next-nearest-neighbor (NNN) interactions exhibits six different phases.\cite{Hikihara2000,Kolezhuk1996} In addition to two N\'eel ordered phases and two phases with a Haldane gap, there exists a large phase region called a gapless chiral phase where the chirality exhibits long range order without accompanying spin order, and a small phase region where there is a gapped chiral phase.
   
\begin{figure}
\includegraphics[width=3in]{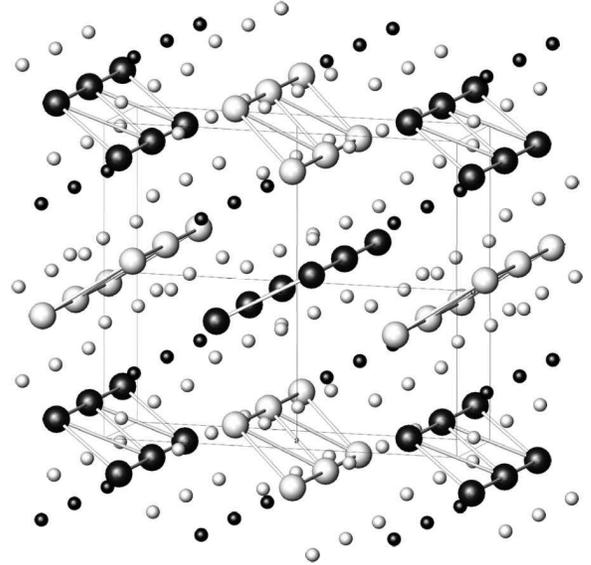}
\caption{The orthorhombic crystal structure of CaV$_2$O$_4$ at room temperature. Large spheres -- two inequivalent V sites, small dark spheres -- Ca, small light spheres -- O.  The crystallographic $a$ and $b$ axes are along the vertical and horizontal directions in the plane of the page, respectively. The $c$ axis is perpendicular to the page. The cuboid denotes the size of a unit cell.}
\label{fig:structure}
\end{figure}

   \cavo\ is a possible candidate for a zig-zag spin $S=1$ chain system.\cite{Kikuchi2001,Fukushima2002} It has an orthorhombic crystal structure (space group \textit{Pnam}) at room temperature as shown in Fig.~\ref{fig:structure}. Vanadium moments at two crystallographically inequivalent sites respectively  form two inequivalent zig-zag spin chains along the $c$-axis. In one of the two chains, the distances between NN and NNN vanadium atoms are 3.01 and 3.08~\AA, respectively, while in the other chain, these two distances are 3.01 and 3.06~\AA, respectively. The smallest interchain vanadium distances are 3.58 and 3.62~\AA.\cite{Niazi2007} Thus one might expect a much smaller interchain coupling as compared to NN and NNN interactions within the chain. 
   
   Previous magnetization and \va\ NMR studies of \cavo\ suggested that the ground state of the system might be a gapless chiral phase.\cite{Fukushima2002,Kikuchi2001}  However,  this finding contradicts earlier powder neutron diffraction studies which showed that the system is in an antiferromagnetic state at 4.2~K\@.\cite{Hastings1967} The neutron measurements indicated a magnetic unit cell in which the $b$ and $c$ lattice constants are doubled and the spin directions in each chemical unit cell are reversed relative to their orientations in neighboring chemical unit cells along the $b$- and $c$-axes. Each chemical unit cell contains 8 vanadium spins and the magnetic spin structure within each chemical unit cell could not be uniquely determined. The neutron diffraction pattern was found to be consistent with three different collinear models with spins parallel to the $b$-axis. By assuming the same spin moment at all vanadium sites, the magnetic moment of each vanadium spin was determined to be $1.06(6)~\mu_{\rm B}$, only about half the value expected for a vanadium spin with $g$-factor $g\approx 2$ and $S=1$.\cite{Hastings1967}
The presence of a low temperature antiferromagnetic phase is also supported by recent magnetization measurements on annealed \cavo\ single crystals, which showed a clear signature of an antiferromagnetic phase  transition at temperature $T_{\rm N}=69$~K\@.\cite{Niazi2007}
   
    In order to obtain further evidence of the magnetic phase transition and to study the magnetic properties in the ordered state, we performed \ox\ and \va\ NMR studies on \ox-enriched powder and single crystal samples, respectively.  The \ox\ NMR spectrum and the nuclear spin-lattice relaxation rate  measurements give strong evidence for a magnetic transition at $T_{\rm N}=78$~K in the powder sample. In contrast to early \va\ NMR measurements,\cite{Kikuchi2001,Fukushima2002} we could not detect a \va\ NMR signal in the powder sample around the normal Larmor frequency in an applied field of 1.67~T in the temperature range of $4.2~\mathrm{K}<T<296$~K\@. Instead, we observed a \textit{zero-field} \va\ NMR signal at $T<45$~K ($f\approx 237$~MHz at 4.2~K).     
    
    The zero-field \va\ NMR signal is observed because of a strong local field at \va\ nuclear sites ($H_{\rm loc}=21.2$~T at 4.2~K) in the ordered state. The local field arises mainly from interaction between nuclei and vanadium core electrons, which are polarized by the ordered 3$d$ electronic spins.\cite{Freeman1963,Carter} This local field points antiparallel to the direction of the local electronic spin moment. By studying how the resonance frequency changes as a function of the direction and magnitude of the applied magnetic field, one can obtain information on  the vanadium spin structure, as will be demonstrated below. We measured the temperature dependence of the ordered moment to study the anisotropy gap of the antiferromagnetic spin wave excitations. We also attempted to measure the temperature dependence of the \va\ nuclear spin-lattice relaxation rate $1/T_1$. However, due to the very broad line and the presence of nuclear quadrupole splitting (the nuclear spin of \va\ is $I = 7/2$), the relaxation curves are highly nonexponential and depend strongly on the saturation condition. Thus, a reliable measurement of the \va\ $1/T_1(T)$ was not achieved.

 The remainder of the paper is organized as follows. Experimental details are explained in Sec.~\ref{sec:exp}. \ox\ and \va\ NMR results are presented in Secs.~\ref{sec:O17} and \ref{sec:V51}, respectively. In Sec.~\ref{sec:summary}, we give a summary of the main results of the paper. 
 
\section{\label{sec:exp}Experimental Details}

Polycrystalline single phase CaV$_2$O$_4$ (sample an-2-116) was synthesized via the solid-state route by reacting V$_2$O$_3$ (99.995\%, MV Labs) with CaO obtained by calcining CaCO$_3$ (99.995\%, Aithaca Chemicals) at 1100~$^{\circ}$C. The chemicals were ground inside a He glove-box, then pressed and sintered at 1200~$^{\circ}$C for 96 hours in flowing 4.5\% H$_2$-He with intermediate grindings. Phase purity was confirmed by powder X-ray diffraction (XRD) on a Rigaku Geigeflex diffractometer using Cu K$\alpha$ radiation in the $2\theta$ range of 10$^\circ$--$90^\circ$.\cite{Niazi2007} 72.1~atomic\% $^{17}$O-enriched oxygen (MSD Isotopes) was used for $^{17}$O-enrichment. About 1~g of the precursor \cavo\ was placed in a Pt foil-lined alumina boat in an evacuated furnace tube, which was then preheated to 750~${^\circ}$C under dynamic vacuum. The pumping line was then closed and the tube backfilled with the $^{17}$O-enriched O$_2$\@. The mass gain on oxidation indicated a nominal composition of CaV$_2$O$_{5.94}$. This was placed in flowing 4.5\% H$_2$-He and reduced as before to CaV$_2$O$_4$ (sample an-2-180E).   The final  $^{17}$O content of the enriched CaV$_2$O$_4$ was about 25\%. Powder XRD was used to confirm that the sample was single phase.

\begin{figure}
	\centering
		\includegraphics[width=3in]{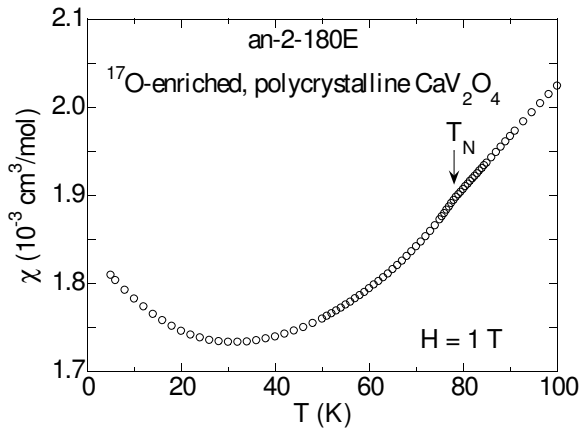}
	\caption{Magnetic susceptibility $\chi$ versus temperature $T$ of the \ox-enriched \cavo\ powder sample measured in a field of $H=1$~T\@. The vertical arrow indicates the position of the antiferromagnetic transition temperature $T_{\rm N} = 78$~K\@.}
	\label{fig:chi}
\end{figure}

 $^{17}$O NMR measurements were performed utilizing a phase-coherent pulse spectrometer in applied fields of 3.0 and 4.7~T\@. The typical $\pi /2$ pulse length is $6\,\mathrm{\mu s}$. The echo signal was produced by a sequence of a $\pi /2$ and a $\pi /3$ pulse,  which produce the maximum echo signal intensity. The separation between these two echo generating pulses was 40~$\mathrm{\mu s}$. The \ox\ NMR spectra were measured by either Fourier transform of half the echo signal or by plotting the area of the echo as a function of the rf frequency (frequency sweep). The nuclear spin-lattice relaxation rates were measured by monitoring the recovery of the echo intensity following a comb sequence of $\pi/2$ saturation pulses. Static magnetization versus temperature was measured in a Quantum Design SQUID magnetometer in a field of $1$~T and in the temperature range 5--100~K to confirm the low temperature magnetic behavior and the ordering temperature $T_{\rm N} = 78$~K\@.  The magnetic susceptibility of the powdered \ox-enriched sample is shown in Fig.~\ref{fig:chi}. The transition temperature is revealed by a small kink in the $\chi(T)$ data at $T_{\rm N}$.

\begin{figure}
	\centering
		\includegraphics[width=3in]{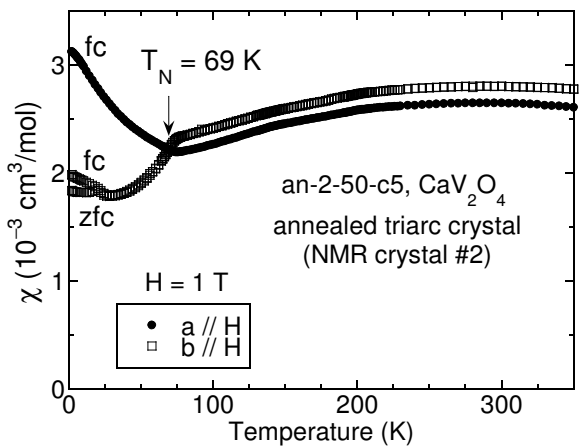}
		\caption{Magnetic susceptibility $\chi$ versus temperature $T$ of \cavo\ crystal \#2 measured with the applied field $H=1$~T along the $a$ and $b$ directions. The vertical arrow indicates the antiferromagnetic transition temperature $T_{\rm N}=69$~K\@. The measurements were carried out under either field-cooled or zero-field-cooled conditions, as indicated.}
	\label{fig:xstal}
\end{figure}

Two \cavo\ crystals were used in \va\ NMR measurements. Crystal \#1 was grown in an optical floating zone furnace while crystal \#2 was grown using a tri-arc crystal pulling method.\cite{Niazi2007,mpc}  The sizes of the crystals \#1 and \#2 are about $5\times5\times10$~mm$^3$ and $1\times1\times2$~mm$^3$, respectively.  Both crystals were annealed at 1200~$^\circ$C under 4.5\%  H$_2$-He flow and the antiferromagnetic phase transition temperatures of the annealed crystals were found from magnetic susceptibility and heat capacity measurements to be 69~K.\cite{Niazi2007} The difference in $T_{\rm N}$ between the powder and single crystal samples may be related to different unit cell volumes of these samples.  The unit cell volumes in the powder and single crystal \cavo\ samples are 296.0(5) and 298.0(5) \AA$^3$, respectively. The reason for this difference is, however, unknown at present. A similar correlation between $T_{\rm N}$ and the lattice volume was observed in EuCu$_2$Ge$_2$.\cite{Hossain2003} The magnetic susceptibility of crystal \#2 is shown in Fig.~\ref{fig:xstal} with the field along $a$ and $b$ directions. The antiferromagnetic transition temperature $T_{\rm N}$ is clearly seen as a bifurcation in the susceptibilities along the two directions.  We note that, when the field is along the $b$ direction, a splitting between zero-field-cooled and field-cooled susceptibility is observed below $T=20$~K\@.  

A search for a zero-field \va\ NMR echo signal was performed at 4.2~K and was found to be located close to a frequency of 237~MHz at that temperature. The echo was produced by a sequence of two pulses with the same pulse length, which was typically 4~$\mu$s and about half the length of a $\pi/2$ pulse. The separation between the pulses was fixed to $18~\mu$s. \va\ NMR spectra were measured by plotting the echo intensity as a function of the magnetic field.  A variable magnetic field from 0 to 2.0~T was produced by an electromagnet. The value of the magnetic field was measured by a Hall magnetometer attached to one of the two magnet pole caps. The difference between the measured field and the field at the position of the sample (measured by the resonance frequency of protons in water) was less than 0.005~T over the whole field range. Individually, the crystals were placed inside copper solenoid coils with the crystal $c$-axis parallel to the coil axis. Measurements of crystal \#1 involved rotation of the field in the $a$-$b$ and $b$-$c$ planes. Measurements of crystal \#2 involved rotation of the field in the $a$-$b$ plane. The rotation of the field was achieved by rotating the cryostat together with the crystal about the crystallographic axes perpendicular to the field plane. The misalignment between the rotation axis and the intended crystallographic axis is estimated to be less than $5^\circ$.

\section{\label{sec:O17}$^{17}$O NMR in Powder Sample of $\mathrm{\bf CaV_2O_4}$}

\begin{figure}
\centering
		\includegraphics[width=3in]{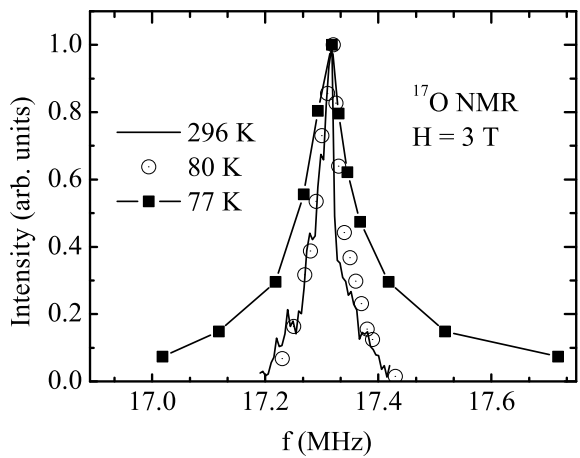}
		\caption{Absorption spectrum of the $^{17}$O NMR signal for an \ox-enriched powder sample of \cavo\ at three different temperatures in an applied magnetic field $H=3.0$~T. A strong inhomogeneous broadening is observed close to the magnetic transition temperature $T_{\rm N}=78$~K\@. The solid line at 77~K is a guide to the eye.}
	\label{fig:O17Spectrum}
\end{figure}

 Figure~\ref{fig:O17Spectrum} displays the \ox\ NMR spectra for the \ox-enriched powder sample of \cavo\  in $H=3$~T at three different temperatures. The spectrum at $T=296$~K was obtained via Fourier transform of half the echo signal while the spectra at $T=80$ and 77~K were  obtained by  frequency sweep. \ox\ nuclei have spin $5/2$ and thus possess a nonzero electric quadrupole moment.  Since the local environments of all oxygen sites do not possess cubic symmetry (point group $m$), one expects a quadrupole splitting of the \ox\ resonance frequencies. We attribute the lack of a powder pattern of the first order quadrupole splitting in the observed spectra to a smaller quadrupole splitting compared to the magnetic broadening of the spectra. The positions of the spectrum peak position show negligible field dependence. The chemical shift $K$ of the peak position is given by $K = A_{\rm hf}\chi$, where $A_{\rm hf}$ is the isotropic hyperfine coupling constant and $\chi$ is the magnetic susceptibility. Using the data in Figs.~\ref{fig:chi} and \ref{fig:O17Spectrum} and additional $\chi(T)$ data (not shown), we obtain the upper limit $A_{\rm hf} \lesssim 1.7~\mathrm{kG}/\mu_{\rm B}$.    
 
 The line gets broader below $T_{\rm N}$ with decreasing temperature.  As shown in Fig.~\ref{fig:O17Spectrum}, the absorption line at $T= 77$~K is much broader than the lines at $T=80$ and 296~K\@.  As will be further shown below, this broadening is a signature of an antiferromagnetic phase transition at $T_{\rm N} =78$~K, where the nuclear spin-lattice relaxation rate $1/T_1$ exhibits a peak.  As the temperature approaches the phase transition temperature, the electronic vanadium spins slow down dramatically and thus induce an inhomogeneous static (on the NMR time scale) dipolar or transfered hyperfine field on neighboring \ox\ sites and broaden the \ox\ NMR line. Due to the fast decrease of signal intensity below $T_{\rm N}$ with decreasing temperature, detailed measurements of the temperature dependence of the line broadening in this temperature range were not pursued.
 
\begin{figure}
	\centering	\includegraphics[width=3in]{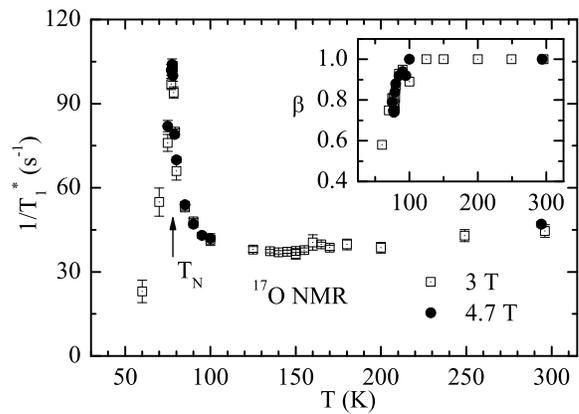}
	\caption{Temperature $T$ dependence of the \ox\ nuclear spin-lattice relaxation rate $1/T_1^*$ [see Eq.~(\ref{eqn:exponential})] of \ox-enriched powder \cavo\ in applied magnetic fields $H=3.0$ and 4.7~T\@.  A strong peak is observed close to the antiferromagnetic transition temperature $T_{\rm N}=78$~K\@, as labeled by the arrow. Inset: the stretching exponent $\beta$ versus $T$\@.}
	\label{fig:O17T1}
\end{figure}

 The  recovery of the \ox\ longitudinal nuclear magnetization $M(t)$ following the saturation pulses is a single exponential function at $T>100$~K. This shows that all the $^{17}$O NMR transitions are saturated and thus there are no satellite peaks located away from the main peaks displayed in Fig.~\ref{fig:O17Spectrum}, consistent with a small quadrupole splitting as discussed above. Below 100~K, deviation from single exponential behavior is observed, indicating a distribution of $1/T_1$ values. In order to extract a characteristic relaxation rate at all temperatures, we fitted the recovery curves  to a stretched exponential function
\begin{equation}
1-\frac{M(t)}{M(\infty )}=\exp \left[-(t/T_1^*)^{\beta}\right].
\label{eqn:exponential}
\end{equation}
  The relation of the parameters $1/T_1^*$ and $\beta$ to the $1/T_1$ distribution underlying a stretched exponential function has been discussed in a recent paper.\cite{Johnston2006} Figure~\ref{fig:O17T1} shows the temperature dependence of $1/T_{1}^*$ and $\beta$ in 
$H=3.0$ and 4.7~T\@. At $T>100$~K, the relaxation rate is almost temperature independent.
Below 100~K, $1/T_1^*$ exhibits a strong enhancement and reaches a peak at $T_{\rm N}=78$~K\@. 
Combining  the above NMR results with magnetization studies of powder and single crystals,\cite{Niazi2007} and with the magnetic susceptibility data in Fig.~\ref{fig:chi}, we identify the $1/T_1^*$ peak temperature as the temperature of an antiferromagnetic phase transition $T_{\rm N}$. 
The peak in the nuclear spin-lattice relaxation rate at $T_{\rm N}$  results from an enhancement and slowing down of the electronic spin fluctuations at wave vectors close to the antiferromagnetic ordering vector as the temperature approaches $T_{\rm N}$ from either side.\cite{Borsa1979}
 
\section{\label{sec:V51}$^{51}$V NMR Below $T_{\rm N}$ in single crystals of $\mathrm{\bf CaV_2O_4}$}

\subsection{Spin Structure at 4.2~K}

   In an external magnetic field $\bm{H}$, the resonance frequency $f$ of the $^{51}$V nuclear spins is given by 
 \begin{equation}
	f = |A \langle \bm{S} \rangle + \bm{H}| \gamma_{\rm v}/2\pi,
	\label{eqn:hyperfine}
\end{equation}  
where $A$ is the hyperfine coupling constant between the nuclear spin and the vanadium electronic spins $\bm{S}$,  $\gamma_{\rm v}$ is the gyromagnetic ratio of $^{51}$V nuclear spins, and $\langle \bm{S} \rangle$ denotes the average electronic spin value in thermal equilibrium. In our experiments, the local field is much larger than the applied field: $A |\langle \bm{S} \rangle| \gg H$. Depending on whether the applied field component along $\langle \bm{S} \rangle$ is parallel or antiparallel to $\langle \bm{S}\rangle$, the resonance frequency shifts to higher or lower values, respectively. If \cavo\ is a collinear antiferromagnet at low temperatures, where opposite spin directions exist, upon application of an external field along the ordering axis, the zero-field spectrum should split into two peaks. These peaks should be symmetrically displaced above and below the zero-field peak frequency.

\begin{figure}
	\centering	\includegraphics[width=3in]{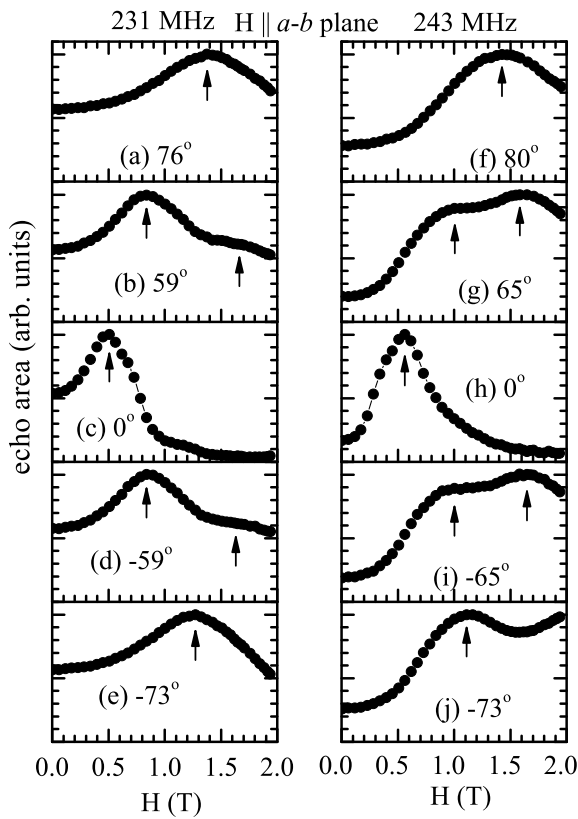}
	\caption{Field swept spectra with the applied magnetic field parallel to the $a$-$b$ plane at rf frequencies of 231~MHz (left panels) and 243~MHz (right panels).  The angles between the field and the average of the two projections of the two spin directions onto the $a$-$b$ plane ($\bm{S}_m^{\prime}$ in Fig.~\ref{fig:spinstrctr}) are labeled in each panel. The arrows indicate the positions of the peaks. $\bm{S}_m^{\prime}$ is approximately parallel to the crystallographic $b$-axis. The spectra were measured at 4.2~K on crystal \#1.}
	\label{fig:specab}
\end{figure}

\begin{figure}
	\centering	\includegraphics[width=3in]{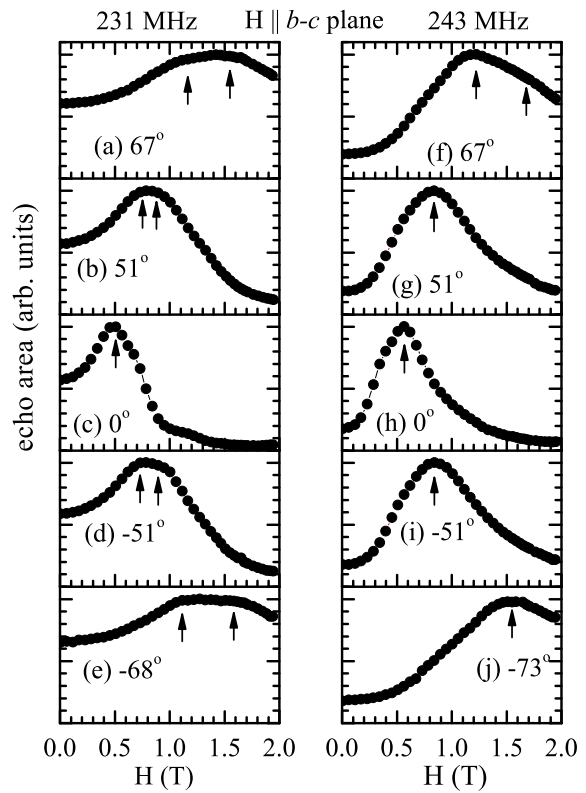}
	\caption{Field swept spectra with the applied magnetic field parallel to the $b$-$c$ plane at rf frequencies of 231~MHz (left panels) and 243~MHz (right panels).  The angles between the field and the average of the two projections of the two spin directions onto the $b$-$c$ plane ($\bm{S}_m^{\prime\prime}$ in Fig.~\ref{fig:spinstrctr}) are labeled in each panel. The arrows indicate the positions of the peaks. The spectra were measured at 4.2~K on crystal \#1.}
	\label{fig:specbc}
\end{figure}

    Figure~\ref{fig:specab} shows the field-swept spectra with the field at different angles in the $a$-$b$ plane, measured with rf frequencies both higher and lower than the zero-field peak frequency $f_0=236.7$~MHz. All measurements in this subsection were performed on \cavo\ crystal \#1. 
    In contrast to a single peak above and a single peak below the zero-field peak frequency $f_0$ when an applied field is present, as expected for a collinear antiferromagnet, instead we see two peaks above $f_0$ and two peaks below $f_0$ in applied fields as shown by the vertical arrows in Fig.~\ref{fig:specab}. Whether each set of two peaks is resolved depends on the angle of the applied field in the $a$-$b$ plane as shown. We infer below that the spectra in Fig.~\ref{fig:specab} (and \ref{fig:specbc}) are consistent with a magnetic structure at 4.2~K that consists of two antiferromagnetic substructures, each of which is a collinear antiferromagnetic arrangement where the angle between the ordering axes of the two substructures is $19(1)^\circ$.  
  The angle labeled in each panel of Fig.~\ref{fig:specab} is the angle between the applied field and the axis $\bm{S}_m^{\prime}$, which is the average of the projections onto the $a$-$b$ plane of the two spin ordering directions (see Fig.~\ref{fig:spinstrctr} below). $\bm{S}_m^{\prime}$ is approximately parallel to the $b$-axis and  is determined by fitting the peak positions versus angle, as will be explained below.
     
     In our discussions of the $^{51}$V NMR results, we assume that the applied magnetic field only shifts the NMR frequency without affecting the electronic spins. In fact, the ordered electronic moments can be tilted by the applied field due to the presence of a torque. However, we can show that the  tilting angle is indeed negligibly small. From the magnetization measurements,\cite{Niazi2007} at 4.2~K,  the susceptibility $\chi$ of single crystal \cavo\ with applied field in the $a$, $b$ or $c$ directions  is $\chi\sim 0.003$~cm$^3$/mol, which corresponds to an induced moment of $0.005~\mu_{\rm B}$ for each vanadium spin in a 2~T field. The tilting angle required to produce such a moment is only $0.3^\circ$, assuming an ordered moment of $1.06~\mu_{\rm B}$.\cite{Hastings1967} Both this angle and the induced moment are negligible to our studies.
      
  The spectra in Fig.~\ref{fig:specab} exhibit a two-peak structure when the field points away from the $\bm{S}_m^\prime$ axis. When measured by field sweep at a fixed frequency $f$, each peak $i\ (=1,2)$ should appear at the field value
\begin{equation}
H_{\mathrm{peak}\,i} = \left|-f_0\cos\alpha_i+\sqrt{f^2-f_0^2\sin^2\alpha_i}\right|/(\gamma_{\rm v}/2\pi),
	\label{eqn:Hpk}
\end{equation}
    where  $\alpha_i$ is the angle between the field and the respective electronic spin ordering direction of a magnetic substructure noted above. 
     The observed two-peak structures for $f>f_0$ and $f<f_0$ strongly indicate the presence of these two different antiferromagnetic spin ordering directions in the system. 
 The ability to resolve the two peaks at the larger angles in Fig.~\ref{fig:specab} (and \ref{fig:specbc}) is related to the larger partial derivative of $|\partial H_{\mathrm {peak}\,i}/\partial\alpha_i|$ of Eq.~(\ref{eqn:Hpk}) at the larger $\alpha_i$ values for $\alpha_i<\pi/2$~rad.
   
\begin{figure}
	\centering	\includegraphics[width=3in]{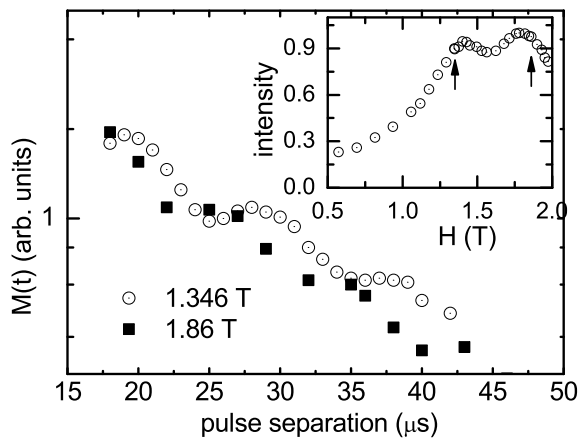}
	\caption{Semilog plot of nuclear spin-spin relaxation curves at external magnetic fields of 1.346 and 1.86~T\@. The errors on the data points are smaller than the sizes of the symbols. The fields are parallel to the $a$-$b$ plane and form an angle of $31^{\circ}$ from the $\bm{S}_m^{\prime}$ direction.  Inset: the field swept spectrum measured at a fixed pulse separation of 18~$\mu$s in the same field direction. The vertical arrows show the field positions of 1.346 and 1.86~T\@. The measurements were performed at 4.2~K on crystal \#2.}
	\label{fig:T2}
\end{figure}

    The two peaks in the same spectrum have different heights, as can be clearly seen in Fig.~\ref{fig:specab} (b), (d), (g), and (i). The reason behind this difference is currently not understood. It may be due to the change of the nuclear spin-spin relaxation times at different field values, since we fix the separation between the two rf pulses for echo generation to be $18~\mu$s.  In fact, we observed a field-dependent oscillation of the spin echo intensity as a function of the separation between the two echo-generating pulses.  Figure~\ref{fig:T2} displays two nuclear spin-spin relaxation curves measured under identical conditions except for different external magnetic fields of 1.346 and 1.86~T, respectively. The oscillation pattern shows a clear field dependence. However, without knowing the detailed functional form of the decay curves, we cannot extrapolate the signal intensity back to zero pulse separation to correct for the nuclear spin-spin relaxation effect. 

     We further note that the different heights of the two peaks cannot be attributed to the different percentage of spins in the two spin substructures. In such a scenario, the spins contributing to the left peaks in Figs.~\ref{fig:specab}(b) and (g) should contribute to the right peaks in Figs.~\ref{fig:specab}(d) and (i) (see Fig.~\ref{fig:pkvsq}), and the left peaks should be higher in one orientation while lower in the other in Figs.~\ref{fig:specab}(b) and (d), and in Figs.~\ref{fig:specab}(g) and (i), respectively. However, the spectra in Figs.~\ref{fig:specab}(b) and (d), and in Figs.~\ref{fig:specab}(g) and (i) are almost the same. The symmetry in the spectra with the field on opposite sides of $\bm{S}_m^{\prime}$, such as in Figs.~\ref{fig:specab}(b) and (d) and in Figs.~\ref{fig:specab}(g) and (i), indicates that the number of spins in the two substructures are the same.

Since the above two-peak structure is observed with the field in the $a$-$b$ plane, these measurements can only detect the difference of the spin projections  of the two substructures onto the $a$-$b$ plane. In order to determine whether or not the projections onto the $b$-$c$ plane  are also different, we also measured the spectra with the field in the $b$-$c$ plane. Some representative spectra with the field in different directions are displayed in Fig.~\ref{fig:specbc}, where the angles listed are described in the caption. As one can see, a two peak structure is still observed when the field is at a large angle from the $\bm{S}_m^{\prime\prime}$ axis. However, the separations between the two peaks are smaller than in Fig.~\ref{fig:specab}, indicating a smaller angle between the two easy axis projections onto the $b$-$c$ plane than onto the $a$-$b$ plane.

\begin{figure}
	\centering	\includegraphics[width=3in]{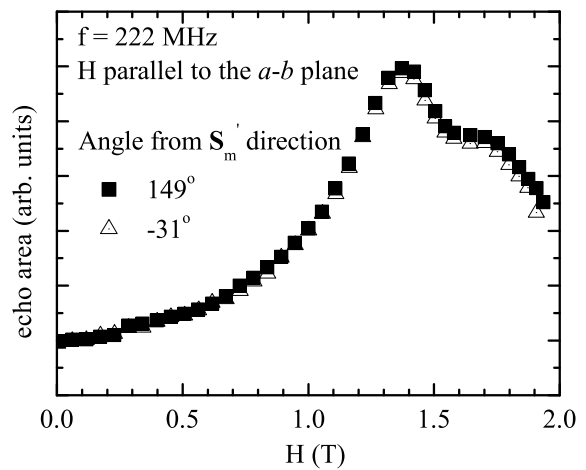}
	\caption{Comparison of two spectra measured at rf frequency $f=222$~MHz with and without the field direction reversed. The fields are parallel to the $a$-$b$ plane and form angles of $-31^{\circ}$ and $149^{\circ}$ from the $\bm{S}_m^{\prime}$ direction, respectively. The spectra were measured at 4.2~K on crystal \#1.}
	\label{fig:fieldinverse}
\end{figure}

   In order to study whether there exists canting and/or an imbalance in the number of spins in opposite directions for each of the two ordered magnetic substructures, we compared the spectra with those measured with the field rotated by $180^\circ$. Figure~\ref{fig:fieldinverse} displays two spectra measured at $f=222$~MHz with the field parallel to the $a$-$b$ plane and $-31^{\circ}$ and $149^{\circ}$ away from the $\bm{S}_m^{\prime}$ direction, respectively. These two spectra are identical within experimental error, indicating the absence of spin canting and the same number of spins in opposite directions within each magnetic substructure.

  Thus we propose a model of the spin structure as shown in Fig.~\ref{fig:spinstrctr}. Various notations used in the model are explained in the caption of Fig.~\ref{fig:spinstrctr}. There are equal numbers of spins in the two antiferromagnetic substructures, each of which consists of collinear antiparallel spins also with equal number. The plane defined by the two ordering directions is parallel neither to the $a$-$b$ nor the $b$-$c$ plane. The average ordered moment direction $\bm{S}_m$ is approximately parallel to the $b$-axis. This is consistent with single crystal anisotropic magnetization measurements versus temperature which showed that below $T_{\rm N}$, the average easy axis of the magnetic structure is approximately the $b$-axis.\cite{Niazi2007}  Note that in our NMR study, we cannot determine the location in the lattice of the two different magnetic substructures.

\begin{figure}
	\centering	\includegraphics[width=3in]{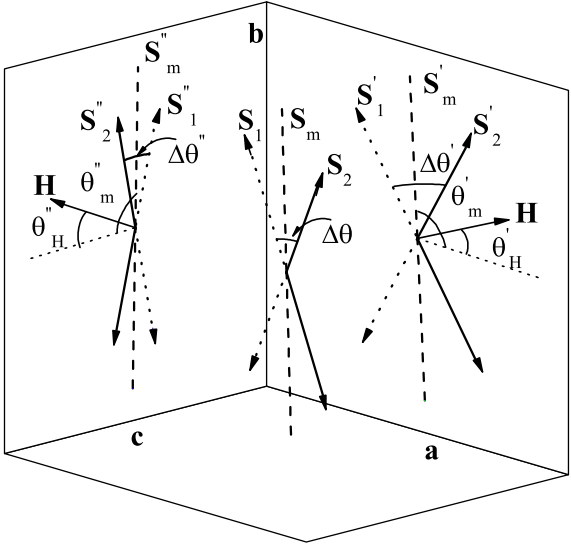}
	\caption{The proposed ordered spin structure in \cavo. There are two different antiferromagnetic ordering substructures with equal numbers of spins, each of which has a collinear antiferromagnetic spin arrangement. $\Delta\theta$, $\Delta\theta^{\prime}$, and $\Delta\theta^{\prime\prime}$ are the angles between these two directions, and their projections on $a$-$b$ and $b$-$c$ planes, respectively. $\bm{S}_m$, $\bm{S}_m^{\prime}$, and $\bm{S}_m^{\prime\prime}$ are the average of the two directions and their projections on $a$-$b$ and $b$-$c$ planes, respectively. $\theta_H^{\prime}$ ($\theta_H^{\prime\prime}$) and $\theta_m^{\prime}$ ($\theta_m^{\prime\prime}$) are the angles formed between a fixed arbitrary experimental reference direction in the $a$-$b$ ($b$-$c$) plane and the applied field $\bm{H}$ and $\bm{S}_m^{\prime}$ ($\bm{S}_m^{\prime\prime}$), respectively. 
}
	\label{fig:spinstrctr}
\end{figure}

   To extract the angle between the spin ordering directions of the two substructures, we measured the dependence of the peak positions versus the field directions at rf frequencies  of 231 and 243~MHz. The results are shown in Fig.~\ref{fig:pkvsq}. For small angles between $\bm{H}$ and $\bm{S}_m^{\prime}$ or $\bm{S}_m^{\prime\prime}$, only one  peak is observed. At larger angles, the positions of two peaks can be resolved. Two different symbols are used to represent the two different spin substructures, while for spectra with single peaks, a third symbol is used. Since the angle between the spin direction and the $b$-axis is much less than one radian,  $\cos\alpha_i$ in Eq.~(\ref{eqn:Hpk}) for the field in the $a$-$b$ plane can be approximated by 
\begin{eqnarray}
\cos \alpha_{1,2}  \approx \cos(\theta_m^{\prime}\pm\Delta\theta^{\prime}/2-\theta_H^{\prime}),
	\label{eqn:Hpkab}
\end{eqnarray}   
where $\alpha_{1,2}$ are the angles between the field and the two spin directions $\bm{S}_1$ and $\bm{S}_2$, respectively.
Similarly, for the field in the $b$-$c$ plane, one has 
\begin{eqnarray}
\cos \alpha_{1,2}  \approx \cos(\theta_m^{\prime\prime}\pm\Delta\theta^{\prime\prime}/2-\theta_H^{\prime\prime}).
	\label{eqn:Hpkbc}
\end{eqnarray}   
 We fitted Eq.~(\ref{eqn:Hpk}) with $\cos\alpha_{1,2}$ given by Eqs.~(\ref{eqn:Hpkab}) and (\ref{eqn:Hpkbc}) to the data in Fig.~\ref{fig:pkvsq}. The free parameters in the fit were $f_0$, $\theta_m^{\prime}$, $\theta_m^{\prime\prime}$, $\Delta \theta^{\prime}$, and $\Delta \theta^{\prime\prime}$. The best fit results are $f_0=236.7(2)$~MHz, $\Delta\theta^{\prime}=18(1)^{\circ}$, and $\Delta\theta^{\prime\prime}=6(1)^\circ$. The fits are shown in Fig.~\ref{fig:pkvsq}. Since the angles between the spins and the $b$-axis are much less than one radian, we have 
\begin{equation}
	\sin\Delta\theta\approx \sqrt{\sin^2\Delta\theta^{\prime}+\sin^2\Delta\theta^{\prime\prime}},
\end{equation}
from which one obtains the angle between the easy axes of the two magnetic substructures to be $\Delta\theta = 19(1)^{\circ}$. 

\begin{figure}
	\centering	\includegraphics[width=2.5in]{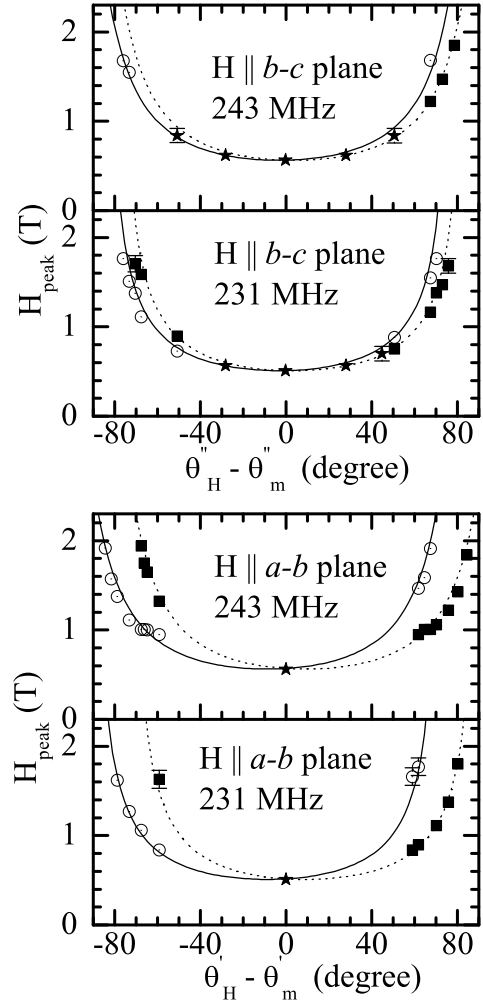}
	\caption{Dependence of the peaks in the spectra at 4.2~K on the direction of the applied magnetic field, with the field in the $b$-$c$ (top two panels) and $a$-$b$ (bottom two panels) planes of crystal \#1, where the rf frequencies are equal to 231 and 243~MHz, respectively. For definitions of the angles $\theta_H^{\prime}$, $\theta_m^{\prime}$, $\theta_H^{\prime\prime}$, and  $\theta_m^{\prime\prime}$, see Fig.~\ref{fig:spinstrctr}. Circles and filled squares correspond to the two different spin ordering directions of the two magnetic substructures, respectively. The symbol $\star$ is used when the two peaks from the two spin directions overlap and only a single peak can be observed. The error in $H_{\rm peak}$ is comparable to the size of the symbols unless shown explicitly. The solid and dotted lines represent the fits by the theoretical prediction in Eq.~(\ref{eqn:Hpk}).}
	\label{fig:pkvsq}
\end{figure}

\begin{figure}
	\centering
\includegraphics[width=3in]{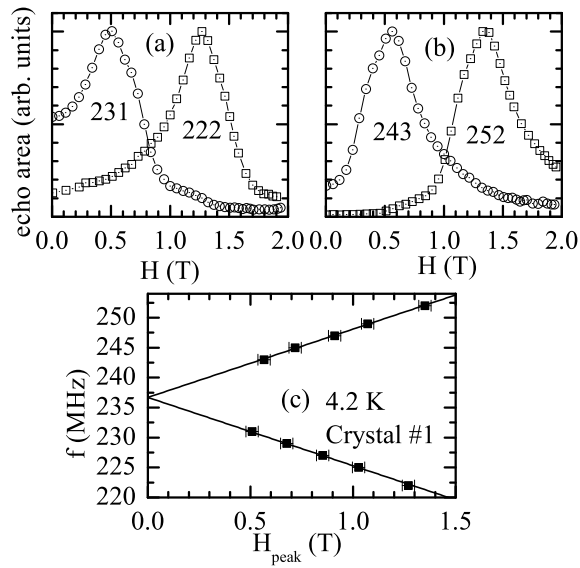}
	\caption{(a) and (b): Field-swept $^{51}$V NMR spectra at four different frequencies at 4.2~K\@. The frequencies are given under each spectrum in units of MHz.  The field is applied parallel to the $\bm{S}_m^{\prime\prime}$ direction. (c): The frequency versus the peak field of the spectra. The solid lines are linear fits by Eq.~(\ref{eqn:linear}). The measurements were done on crystal \#1 at 4.2~K\@.}
	\label{fig:spechparb}
\end{figure}

   In addition to the study of the angular dependence of the peak positions, we also measured their frequency dependences to further confirm the proposed spin structure. Figures~\ref{fig:spechparb}(a) and (b) show the field swept spectra with the field $\bm{H}$ pointing along the $\bm{S}_m^{\prime\prime}$ direction, at rf frequencies lower and higher than $f_0=236.7$~MHz, respectively.  Note that when $\bm{H}\parallel \bm{S}_m^{\prime\prime}$, the two magnetic substructures have the same peak positions of the spectra (see the zero-angle data in Fig.~\ref{fig:pkvsq}). The peaks in Figs.~\ref{fig:spechparb}(a) and (b) both shift to higher fields when the frequency shifts further away from $f_0$, respectively. The peak positions $H_{\rm peak}$ versus rf frequency are plotted in Fig.~\ref{fig:spechparb}(c). The two sets of data points can be well fitted by the two linear equations
\begin{equation}
	f = f_0 \pm H_{\rm peak} \gamma/2\pi,
	\label{eqn:linear}
\end{equation}
where $f_0$ is the peak frequency of the spectrum at zero applied field.
   A fit of Eq.~(\ref{eqn:linear}) to the data gives $\gamma/2\pi=11.4(2)$~MHz/T, and $f_0= 236.7(1)$~MHz. Assuming $\bm{S}_m$ to be parallel to the $b$-axis,  the value of $\gamma$ should be $\gamma/2\pi=(\gamma_{\rm v}/2\pi)\cos(\Delta\theta/2) = 11.07$~MHz/T, where $\Delta\theta=19^\circ$ is the above angle between the ordering directions of the two magnetic substructures. This value of $\gamma/2\pi$ is very close to the above fitting result. 
   
\subsection{Ordered Moment and Its Temperature Dependence}

  In this subsection, we will study the temperature dependence of the vanadium ordered moment, which provides evidence for an energy gap in the antiferromagnetic spin wave excitation spectrum, arising from anisotropy effects. Then we will discuss the value of the saturation vanadium spin moment at low temperatures.  Measurements in this subsection were performed on crystal \#2. The experiment was set up to allow field rotation in the $a$-$b$ plane. By rotating the field in the $a$-$b$ plane, the $\bm{S}_m^{\prime}$ direction (see Fig.~\ref{fig:spinstrctr}) is identified as the direction along which the peak position in the spectrum is a minimum at a fixed rf frequency away from $f_0$ (see Fig.~\ref{fig:pkvsq}). After identifying the $\bm{S}_m^{\prime}$ direction, all subsequent measurements of the spectra were performed versus $H$ at fixed rf frequencies with the field along the $\bm{S}_m^{\prime}$ direction. 
  
\begin{figure}
	\centering
\includegraphics[width=2in]{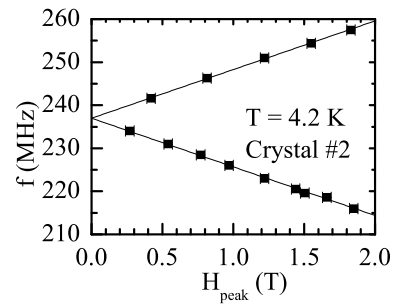}
	\caption{Rf frequency $f$ versus peak $H_{\rm peak}$ in field swept spectrum in crystal \#2 at 4.2~K\@. The field is applied along the $S_m^{\prime}$ direction. The solid lines are fits with Eq.~(\ref{eqn:linear}).}
	\label{fig:fvsHpk2}
\end{figure}

  With the field along $\bm{S}_m^{\prime}$, only a single peak is observed in the spectrum at each frequency (see Fig.~\ref{fig:spechparb}). In Fig.~\ref{fig:fvsHpk2}, we display the frequency dependence of the peak position at 4.2~K\@. Similar to the case of crystal \#1 (Fig.~\ref{fig:spechparb}), the data points can be well fitted by two straight lines. A fit of Eq.~(\ref{eqn:linear}) to the data gives $f_0=236.98(8)$~MHz and $\gamma/2\pi=11.3(1)$~MHz/T\@. This value of $\gamma$ is in agreement with the fitting value in crystal \#1. However, the value of $f_0$ is slightly larger than in crystal \#1. This slight difference may be due to sample-dependent differences. 
  
\begin{figure}
	\centering
\includegraphics[width=3in]{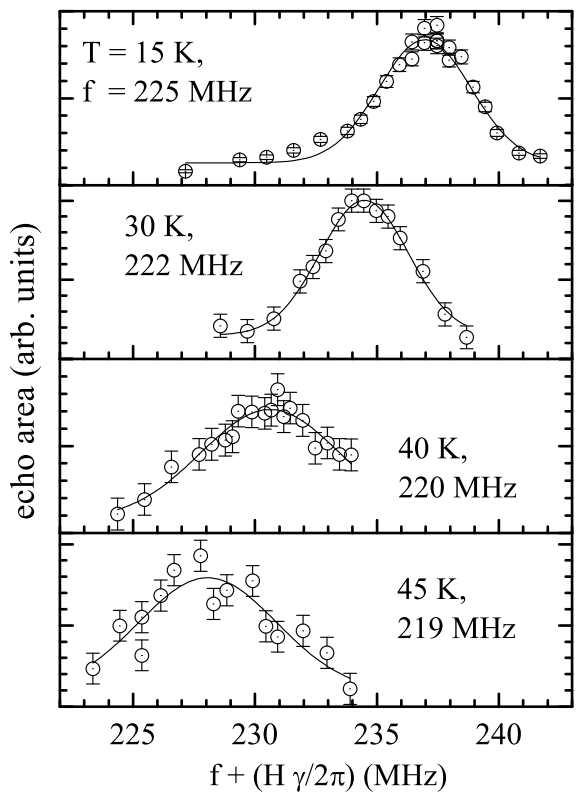}
	\caption{Field swept spectra at different temperatures on crystal \#2. The temperature and the rf frequency for each measurement are labeled in each panel. The solid lines are fits by Eq.~(\ref{eqn:gauss}) to extract the peak positions.}
	\label{fig:spectraT}
\end{figure}

  Figure~\ref{fig:spectraT} displays representative spectra measured at four different temperatures. For comparison between the different spectra, the $x$-axis has been converted to the quantity $f+H\gamma/2\pi$, with $\gamma/2\pi=11.3$~MHz/T\@. As the temperature increases, the signal intensity decreases rapidly and the spectra can only be measured below 45~K\@. In order to more accurately determine the peak position of the spectra, we fitted the original field swept spectra (with the $x$-axis being $H$) by a Gaussian function
\begin{equation}
I(H)= A+B\exp[-2(H-H_{\rm peak})^2/W^2],
\label{eqn:gauss}	
\end{equation}
with $A$, $B$, $H_{\rm peak}$, and $W$ as fitting parameters.
 The zero field peak frequency $f_0$ is then determined from
 \[
 f_0=f\pm H_{\rm peak}\gamma/2\pi,
 \]
  where $\gamma/2\pi=11.3$~MHz/T and the $+$ and $-$ signs correspond to the cases of $f<f_0$ and $f>f_0$, respectively. In order to determine whether $f<f_0$ or $f>f_0$, spectra were measured with at least two different frequencies at each temperature. With the correct choices of the $+$ or $-$ signs, the obtained $f_0$ values for different spectra as in Fig.~\ref{fig:spectraT} are the same within experimental error at each temperature. The final $f_0$ value is an average over all calculated $f_0$ values for various spectra at the same given temperature.      

\begin{figure}
	\centering
\includegraphics[width=3in]{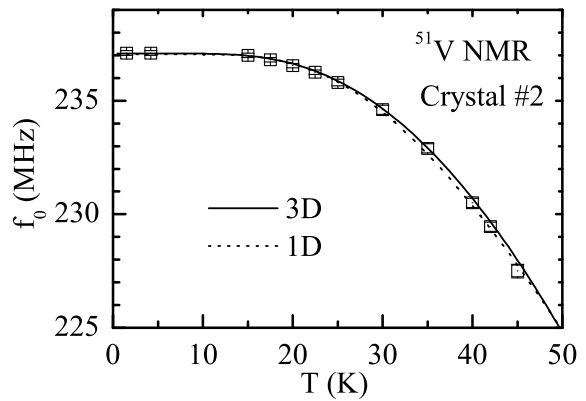}
	\caption{Temperature dependence of the $^{51}$V NMR spectra peak position in zero applied field in \cavo\ crystal \#2. The dotted and solid curves are fits by Eqs.~(\ref{eqn:MTfit1}) with one-dimensional spin wave dispersion and (\ref{eqn:MTfit3}) with three-dimensional dispersion, respectively.}
	\label{fig:V51fvsT}
\end{figure}

 Figure~\ref{fig:V51fvsT} shows the temperature dependence of  $f_0$.  Since the temperature dependence of the hyperfine coupling constant can be ignored,\cite{Turov}  $f_0(T)$ is directly proportional to the ordered local moment.  The ordered moment is almost temperature independent at $T<15$~K. In the spin wave theory of a three-dimensional antiferromagnet \textit{without anisotropy}, the initial decrease of the ordered moment with temperature $T$ should follow a $T^2$ dependence.\cite{Kubo1952} Fitting the data by a power law gives an exponent $\ge 3.5$, an unphysically large value (not shown). The temperature independence below $T<15$~K thus indicates the presence of an anisotropy-induced energy gap for spin wave excitations.\cite{Jaccarino1965} 

Before estimating the gap energy, we will first estimate the value of the saturation  moment at $T=0$~K\@. The local moments reach their saturation value at $T<15$~K\@. From the value of $f_0=237$~MHz at $T<15$~K, one obtains a local field value of $H_{\rm loc}= f_0/(\gamma_{\rm v}/2\pi) =21.2$~T\@. In order to infer the value of local moment from the local field value, the hyperfine coupling constant $A$ and the $g$-factor have to be determined. 
 With known values of $A$ and $g$, the ordered moment $\langle \mu_z \rangle$ is  $|\langle \mu_z \rangle| = g\mu_{\rm B}H_{\rm loc}/A$.  The local field is dominated by the contact interaction through the polarized core electrons, which is approximately proportional to the number of unpaired electronic spins in the $3d$ orbitals with a proportionality constant of 12.5~T per unpaired electron to within 20\%.\cite{Carter}  However, beside the contact interaction, orbital effects\cite{Freeman1963} and transfered hyperfine coupling with the neighboring V$^{3+}$ ions\cite{Kikuchi1996}  may also contribute significantly to the local field at the $^{51}$V nuclear site.  
 
 In the absence of a knowledge of the orbital effects and transfered interactions, we will estimate a possible range of the $A/(g\mu_{\rm B})$ value in \cavo\ using the known values of $A/(g\mu_{\rm B})$ in other V$^{3+}$ compounds.  The values of $A/(g\mu_{\rm B})$ in LaVO$_3$, YVO$_3$, and V$_2$O$_3$ are $16.8$, $20.8$, and $15.5$~T/$\mu_{\rm B}$, respectively.\cite{Kikuchi1994} For V$^{3+}$ in Al$_2$O$_3$, EPR measurements gave $A/(g\mu_{\rm B})=13.3$~T/$\mu_{\rm B}$.\cite{Zverev1960} The range of $A/(g\mu_{\rm B})$ in the above four compounds is between $13.3$ and $20.8$~T/$\mu_{\rm B}$.  Assuming $A/(g\mu_{\rm B})$ in \cavo\ lies in the same range, the low temperature ordered vanadium moment in \cavo\ is thus in the range of 1.02--1.59~$\mu_{\rm B}$. This $|\langle \mu_z\rangle|$ range is consistent with the value $1.06(6)~\mu_{\rm B}$ extracted from the previous neutron powder diffraction study.\cite{Hastings1967} 

\begin{figure}
	\centering
		\includegraphics{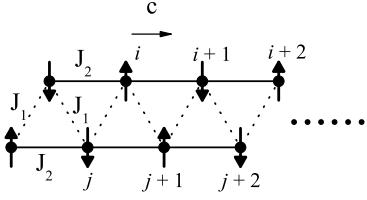}
	\caption{Zig-zag spin structure in CaV$_2$O$_4$.  Due to the alternation of the spin directions along the $c$-aixs, the interaction between the spins in the two legs  of the zig-zag chain are essentially decoupled. The possible misalignment of $19^{\circ}$ between spins in the two legs of the zig-zag chain is ignored. $J_1$ and $J_2$ are the nearest-neighbor inter and intra-leg exchange interactions, respectively.}
	\label{fig:zigzag}
\end{figure}

Next we estimate the energy gap for the spin wave excitations. From the neutron diffraction studies,\cite{Hastings1967} we know that the spins reverse their ordering direction upon moving along the $c$-axis. Therefore the spin structure in the ordered state within a zig-zag chain should look as in Fig.~\ref{fig:zigzag}, where  we ignore the possible small misalignment of $19^{\circ}$ between spins in the two magnetic substructures discussed above. Because each spin in one leg of the chain couples by the same exchange constant $J_1$ to two spins in the other leg that are ordered in opposite directions, we expect the effective coupling between the two legs within a zig-zag chain is greatly reduced in the ordered state. As a result, we will consider the nearest-neighbor interactions within the leg $J_2$ as the only dominant magnetic interaction and treat the effect of interleg interaction within a zig-zag chain  as a weak interchain interaction. For simplicity, we will use a single exchange constant $J^{\prime}$ to characterize the effect of the interchain interactions. To include the effect of anisotropy, we assume a single ion anisotropy in the system with a single direction of easy axis. Then, the Hamiltonian can be written as
\begin{equation}
	H = \sum_{i} \left(2 J_2 \bm{S}_i\cdot \bm{S}_{i+1} -  \frac{1}{2}K S_{iz}^2\right)+2\sum_{\langle i,j\rangle} J^{\prime} \bm{S}_i\cdot \bm{S}_{j+1},
	\label{eqn:simpleH}
\end{equation}
where $K$ is the anisotropy constant, the index $i$ runs through the spins in one leg of the chain, and the summation $\langle i,j\rangle$ runs through all interleg and interchain nearest-neighbor pairs. 

 The Fourier transform of the exchange interactions is
\[
\mathcal{J}(q)=\sum_jJ(\bm{r}_{ij})\exp(-iq\cdot\bm{r}_{ij}),
\]
 where $\bm{r}_{ij}$ connects two spins in opposite sublattices, $J(\bm{r}_{ij})>0$ is the nearest-neighbor exchange constant, and the index $j$ runs through all the nearest-neighbor spins of spin $i$ in the opposite sublattice (each sublattice consists of spins in the same direction). The spin wave dispersion relation is given  by\cite{Lovesey}
\begin{equation}
	E_q = \left\{[2S\mathcal{J}(0)+ KS]^2 - [2S\mathcal{J}(q)]^2\right\}^{1/2},
	\label{eqn:disp}
\end{equation}
 where we ignored interactions between spins in the same sublattice. The spin wave gap value is given by the value of $E_q$ at $q=0$. In the limit of small anisotropy $K \ll 2\mathcal{J}(0)\approx 4J_2$, the gap energy is given by 
\begin{equation}
E_G	= 2S[\mathcal{J}(0)K]^{1/2}.
\label{eqn:EG}
\end{equation}

In the spin wave theory, the decrease of sublattice magnetization is due to the thermal activation of spin wave excitations. In the above bipartite antiferromagnetic system,\cite{Lovesey}
\begin{equation}
	\langle S_z(0)\rangle -	\langle S_z(T)\rangle = \frac{V}{(2\pi)^3}\int\langle n_q\rangle \frac{2\mathcal{J}(0)S+KS}{E_q} d^3q,
	\label{eqn:integral}
\end{equation}
where the integral is limited to the first Brillouin zone of one sublattice, $V$ is the sample volume per sublattice site,  and 
\[
\langle n_q\rangle = \frac{1}{e^{E_q/k_{\rm B}T}-1}
\]
is the number of thermally excited antiferromagnetic magnons.

  The dispersion relation in Eq.~(\ref{eqn:disp}) depends on the spin lattice structure and the exchange interactions $J(\bm{r})$. For a quasi one-dimensional chain with interchain coupling $J^{\prime}$, at temperatures $ T \gg J^{\prime}/k_{\rm B} $, one can ignore the dispersion perpendicular to the chain direction. 
 Then for small values of $q_c$, which is the $\bm{q}$ vector component along the chain, one can perform a Taylor series expansion of $|\mathcal{J}(q)|^2$ as
\begin{equation}
	|\mathcal{J}(q)|^2\approx |\mathcal{J}(0)|^2\left[1- l^2q_c^2\right],
	\label{eqn:smallk}
\end{equation}
where $l$ is the nearest-neighbor distance within the leg. 
 
  At $T\ll J_2/k_{\rm B}$, only spin waves at small $q_c$ values have significant contributions to the integral in Eq.~(\ref{eqn:integral}), so one can change the limits of integral for $q_c$ in Eq.~(\ref{eqn:integral}) to $\pm \infty$.  The small $q$ approximation is valid only at temperatures where $1-	|\langle S_z(T)\rangle |/|\langle S_z(0)\rangle|<0.1$,\cite{Jaccarino1965} which is satisfied within our experimental temperature range. 
Substituting Eq.~(\ref{eqn:smallk}) into Eq.~(\ref{eqn:disp}), and changing the limits of integral for $q_c$ in Eq.~(\ref{eqn:integral}) to $\pm \infty$, one obtains in the limit of small anisotropy $K\ll 2\mathcal{J}(0)$ and $T\ll E_G$
\begin{equation}
		1-	\frac{\langle S_z(T)\rangle }{\langle S_z(0)\rangle} \approx B e^{-E_G/k_{\rm B}T}(E_G/k_{\rm B}T)^{-1/2},
		\label{eqn:ST1d}
\end{equation}
where 
\begin{equation}
B\approx\frac{2}{\sqrt{2\pi}} = 0.80.
\label{eqn:B1}
\end{equation}

Equation~(\ref{eqn:ST1d}) is valid at temperatures $J^{\prime}/k_{\rm B}\ll T\ll J_2/k_{\rm B}$. In \cavo, $J^\prime/k_{\rm B}$ might fall within the experimental temperature range in Fig.~\ref{fig:V51fvsT} ($1.5\le T\le45$~K). So it is useful to consider the other limit of $T\ll J^{\prime}/k_{\rm B} \ll J_2/k_{\rm B}$, where a three dimensional dispersion is more appropriate. Applying a small $q$ approximation, one obtains
\begin{equation}
	|\mathcal{J}(q)|^2\approx |\mathcal{J}(0)|^2\left\{1-\eta^2(V/2)^{2/3}\left[q_c^2+j^{\prime} (q_a^2+q_b^2)\right]\right\},
	\label{eqn:smallk3}
\end{equation}
where for simplicity, we assumed an isotropic dispersion in the $a$-$b$ plane, $\eta$ is a geometrical factor of order one which depends on the spin structure,\cite{Moriya1956} and $j^{\prime}$ is of the order of $J^\prime/J_2$.
By combining Eqs.~(\ref{eqn:disp}), (\ref{eqn:integral}), and (\ref{eqn:smallk3}), and changing the three limits of integrations to $\pm\infty$ in Eq.~(\ref{eqn:integral}), then instead of Eq.~(\ref{eqn:ST1d}), we have in the limit of small anisotropy $K\ll 2\mathcal{J}(0)$ and $T\ll E_G$\cite{Jaccarino1965}
\begin{equation}
		1-	\frac{\langle S_z(T)\rangle }{\langle S_z(0)\rangle} \approx Be^{-E_G/k_{\rm B}T}(E_G/k_{\rm B}T)^{-3/2},
		\label{eqn:ST3d}
\end{equation}
where
\begin{equation}
B\approx \frac{\sqrt{2}\alpha}{\pi^\frac{3}{2}\eta^3j^{\prime}}
\label{eqn:B3}
\end{equation}
with $\alpha\equiv K/[2\mathcal{J}(0)]$.

     Using Eqs.~(\ref{eqn:ST1d}) and (\ref{eqn:ST3d}) and the relation
\[
 f_0(T)/f_0(0) = |\langle S_z(T)\rangle|/|\langle S_z(0)\rangle|,
\]
one obtains the variation of the zero-field \va\ NMR resonance frequency $f_0$ as
(for 1D)
\begin{equation}
f_0(T)=f_0(0)\left[1-B e^{-E_G/k_{\rm B}T}(E_G/k_{\rm B}T)^{-1/2}\right]
\label{eqn:MTfit1}
\end{equation}
or (for 3D)
\begin{equation}
f_0(T)=f_0(0)\left[1-B e^{-E_G/k_{\rm B}T}(E_G/k_{\rm B}T)^{-3/2}\right],
\label{eqn:MTfit3}
\end{equation}
depending on whether a one-dimensional (1D) or three-dimensional (3D) dispersion is used for $E_q$. 
We fitted Eqs.~(\ref{eqn:MTfit1}) and (\ref{eqn:MTfit3}) to the $f_0(T)$ versus $T$ data in Fig.~\ref{fig:V51fvsT} at $T\leq 45$~K with $f_0(0)$, $B$, and $E_G$ as free parameters.  The best fit results are $f_0(0)=237.04(5)$~MHz, $E_G=98(5)$~K, $B=0.51(7)$ for the 1D dispersion with Eq.~(\ref{eqn:MTfit1}), and $f_0(0)=237.08(6)$~MHz, $E_G=64(5)$~K, and $B=0.27(6)$ for the 3D dispersion with Eq.~(\ref{eqn:MTfit3}). The best fit curves are shown as the dotted and solid curves in Fig.~\ref{fig:V51fvsT}, respectively. Since Eqs.~(\ref{eqn:MTfit1}) and (\ref{eqn:MTfit3}) are derived under the two limiting conditions of $T\gg J^{\prime}/k_{\rm B}$ and $T\ll J^{\prime}/k_{\rm B}$, respectively, one can expect that the actual $E_G$ value might lie somewhere between 64 and 98~K\@.  Given a value of $E_G$, we can make a rough estimate of the anisotropy constant $K$. From the magnetization study, one estimates that the intrachain nearest-neighbor exchange constant to be of the order of $J_2/k_{\rm B}\sim 200$~K\@.\cite{Niazi2007} Taking $S=1$, $E_G=81$~K, and $\mathcal{J}(0)\approx2J_2$, we thus have from Eq.~(\ref{eqn:EG}) that $K/k_{\rm B}\sim4$~K\@. 
  
  The above fitting value of $B=0.51(7)$ from 1D dispersion is similar to the calculated value of 0.80.  The fitting value of $B=0.27(6)$ from the 3D dispersion constrains the value of $J^\prime$ in the 3D model. Taking  $\alpha \approx K/4J_2 \sim 0.01$, $\eta\sim1$, and $B=0.27$, then from Eq.~(\ref{eqn:B3}) one has $j^{\prime}\sim 0.01$ and $J^{\prime}/k_{\rm B}\sim j^{\prime}J_2/k_{\rm B}\sim 2$~K\@. However, this value of $J^\prime$ seems inconsistent with the initial assumption  of $T\ll J^{\prime}/k_{\rm B}$ required for the three-dimensional model to be valid. Therefore, Eq.~(\ref{eqn:MTfit1}) of the 1D model  might provide a better approximation to the $f_0(T)$ data.
  
\section{\label{sec:summary}Summary and Conclusions}

We have presented \ox\ and \va\ NMR results on the zig-zag spin chain compound CaV$_2$O$_4$. The strong inhomogeneous broadening and a peak in the nuclear spin-lattice relaxation rate versus temperature of \ox\ NMR confirm the presence of an antiferromagnetic phase transition at 78~K in a powder sample. The crystals we studied have $T_{\rm N}=69$~K\@. \va\ NMR in the ordered state of crystals reveals the presence of two antiferromagnetic substructures at 4.2~K, each of which is collinear and which form an angle of $19(1)^{\circ}$ between them with the average direction approximately parallel to the $b$-axis. The origin and location in the lattice of the different spin substructures remain unknown. However, we speculate that the two magnetic substructures are associated with the two inequivalent V$^{3+}$ $S=1$ zig-zag spin chains in the orthorhombic crystal structure, respectively. Magnetic neutron diffraction studies are required to further characterize the magnetic structure below $T_{\rm N}$. The temperature dependence of the zero-field resonance frequency at low temperatures suggests the presence of magnetic anisotropy and an energy gap in the spin wave excitation spectrum.  The energy gap is estimated from spin wave theory to be between 64 and 98~K\@.

\acknowledgments{We acknowledge F. Borsa, B. Lake, and A. Kreyssig for useful discussions. Work at the Ames Laboratory was supported by the Department of Energy-Basic Energy Sciences under Contract No.\ DE-AC02-07CH11358.}

\end{document}